**Author:**

Dr. Claudio Maccone

**International Academy of Astronautics (IAA) and Istituto Nazionale di Astrofisica (INAF), Italy,
clmaccon@libero.it**

# LIFE EVOLUTION STATISTICS (Evo-SETI) COLLECTED IN A NEW BOOK

**Abstract.** In the past ten years (2010-2020) this author published some 15 highly mathematical papers in prestigious journal like Life, the International Journal of Astrobiology and Acta Astronautica, about his new Evo-SETI (Evolution and SETI) Theory.

Particularly in Refs. [1] thru [4]) it is proven that key features of Evo-SETI are:

1) The Statistical Drake Equation is the extension of the classical Drake equation into Statistics. Probability distributions of the number of ET civilizations in the Galaxy (lognormals) are given, and so is the probable distribution of the distance of ETs from us.
2) Darwinian Evolution is re-defined as a Geometric Brownian Motion (GBM) in the number of living species on Earth over the last 3.5 billion years. Its mean value grows exponentially in time and Mass Extinctions of the past are accounted for as unpredictable low GBM values.
3) The exponential growth of the number of species during Evolution is the geometric locus of the peaks of a one-parameter family of lognormal distributions (b-lognormals, starting each at a different time b=birth) constrained between the time axis and the exponential mean value. This accounts for cladistics (i.e. Evolution lineages).
4) The lifespan of a living being, let it be a cell, an animal, a human, a historic human society, or even an ET society, is mathematically described as a finite b-lognormal. This author then described mathematically the historical development of eight human historic civilizations, from Ancient Greece to the USA, by virtue of b-lognormals.
5) Finally, the b-lognormal's entropy is the measure of a civilization's advancement level. By measuring the entropy difference between Aztecs and Spaniards in 1519, this author was able to account mathematically for the 20-million-Aztecs defeat by a few thousand Spaniards, due to the latter's technological (i.e. entropic) superiority. The same might unfortunately happen to Humans when they will face an ET superior civilization for the first time.

Now the question is: whenever a new exoplanet is discovered, where does that exoplanet stand in its evolution towards life as we have it on Earth nowadays, or beyond? This is the central question of SETI. In this paper we provide mathematical criteria to answer this question within the framework of the Evo-SETI Theory, thus creating the EVO-SETI SCALE. We are now glad to inform our readers that all these mathematical papers have been collected into a single book about Evo-SETI Theory to be published within the year 2020 by Springer. The Front Cover appears here below, as per the Springer website: https://www.springer.com/gp/book/9783030519308.

# Claudio Maccone

# Evo-SETI

## Life Evolution Statistics on Earth and Exoplanets

Springer

# Life Evolution Statistics on Earth and Exoplanets

This book is about the Evolution of Life on Earth as well as on Exoplanets, where Alien Civilizations might live. SETI, the Search for ExtraTerrestrial Intelligence, has been trying to detect Alien Civilizations scientifically since 1960. Evo-SETI (Evolution and SETI) is thus the appropriate title for this book, merging Evolution and SETI in a strong mathematical framework. But fear not if this book is "too mathematical"! We'll take you step-by-step to understand it. And it all started back in the 1980's, when this author watched for the first time the TV series "*Life on Earth: A Natural History*" by David Attenborough https://en.wikipedia.org/wiki/Life_on_Earth_(TV_series). At that time, this author was a mathematical physicist in his forties, and he then decided to cast the evolution of life on Earth into mathematical equations. It took him over thirty years to create the mathematical book that you now see here.

1. **OVERCOME Theorem (formerly Peak-Locus Theorem): growth of Life on Earth over the last 3.5 billion years.**

We shall overcome…we shall overcome… https://www.youtube.com/watch?v=RkNsEH1GD7Q
This famous pacifist song by Joan Baez inspired the author (now, in November 2019) to rename "OVERCOME Theorem" the following mathematical result, that was his most important mathematical discovery made during the ten years 2010-2020. But he used to call it "Peak-Locus Theorem" in all his papers and books published in between 2010 and 2020. The older name "Peak-Locus Theorem" is mathematically more correct, since it refers to the geometric locus of the peaks, as shown by the red solid curve in Figure 1 hereafter. But the name "Peak-Locus Theorem" also somehow hides the intuitive ideas that we now explain in a popular way for the reader's benefit.

The significance of this theorem may be immediately understood easily. Please have a look at Figure 1.

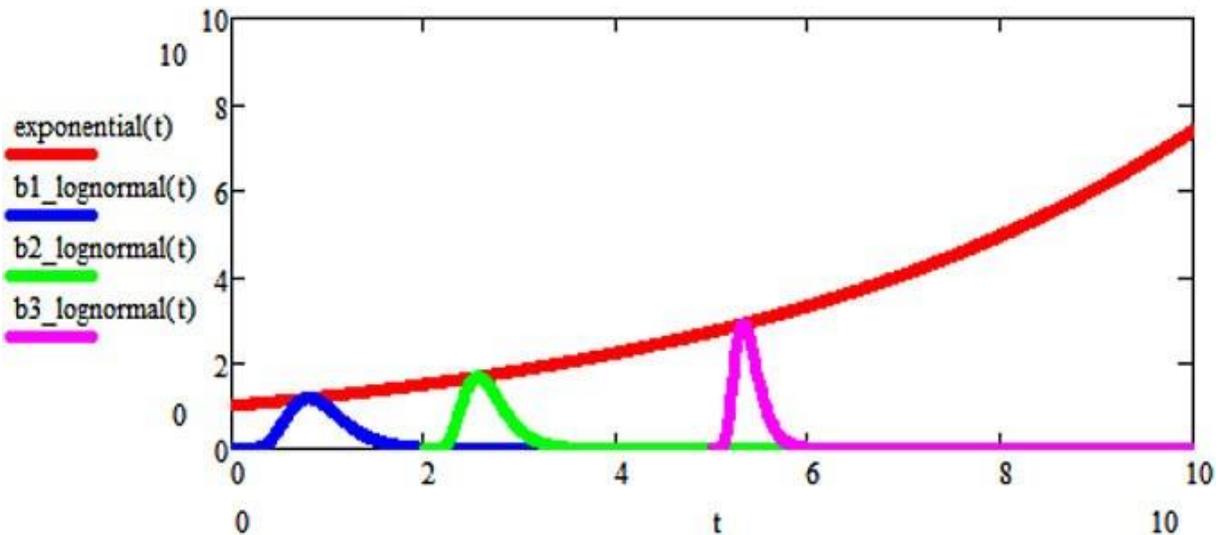

**Figure 1. OVERCOME Theorem (previously called PEAK-LOCUS Theorem (PLT) in all papers, 2010-2020).**

On the horizontal axis is the time, ranging, say, from 0 to 10 in some arbitrary time units.
On the vertical axis the solid red curve is an exponential, ranging from 1 (since $e^{constant*0}=1$) to a value of about 7.6.
In between the horizontal time axis and the increasing red exponential are three different curves, named b1_lognormal (the blue curve), b2_lognormal (the green curve) and b3_lognormal (the fuchsia curve). Please worry not if you don't understand the meaning of the "b-lognormal" name: you'll do so later. But just look at their shape: they start each at a different instant in time: b1=0 (meaning that 0 is the birth time for the b1_lognormal), b2=2 (meaning birth time 2 for the b2_lognormal) and b3=5 (meaning birth time 5 for the b3_lognormal). Then, going left to right, each curve elongates itself in order to have its maximum exactly on the red exponential curve. But… while doing so, the peak of each curve becomes narrower and narrower, in such a way that *the area under the curve is exactly the same for all curves* and, conventionally, we assume that the common value of this *area equals one.*
So the b-lognormals both stretch up and slim up at the same time, because the area under them is assumed to be the same for all of them (that is called the "normalization condition" fulfilled by the b-lognormals).

How does all this apply to reality ?
Well, think of Evolution of Life on Earth over the last 3.5 billion years:
1) For simplicity, suppose that life started exactly 3.5 billion years ago with the first molecule capable of reproducing itself: RNA (i.e. the "RNA world", then leading to the "DNA world" still dominating life today).
2) Then suppose that *each b-lognormal corresponds to a certain Living SPECIES*, born during the course of Evolution just at the instant where that b-lognormal starts.
3) Clearly, the number of Living Species grew up enormously in the 3.5 billion years of Evolution: from 1 (RNA, just 3.5 billion years ago) to 50 million nowadays, as most biologists say. In other words, we now assume that the vertical axis in Figure 1 represents the Number of Species living on Earth while the time elapsed over 3.5. billion years.
4) Is this realistic? In other words, is it realistic to assume that the red upper curve is exactly an exponential? Clearly not so, since we know that many Species died in the course of Evolution, like dinosaurs, killed by an asteroid that hit the Earth about 65 million years ago. So, the "exact" exponential must be replaced by something else: a curve that we don't know well, since it actually may have oscillated up and down around the exponential in an unpredictable way.
5) No problem: mathematicians have a way to get around these difficulties. They replace the exponential curve by a "stochastic process", i.e. a fluctuating curve just like the two different curves shown in the following Figure 2.

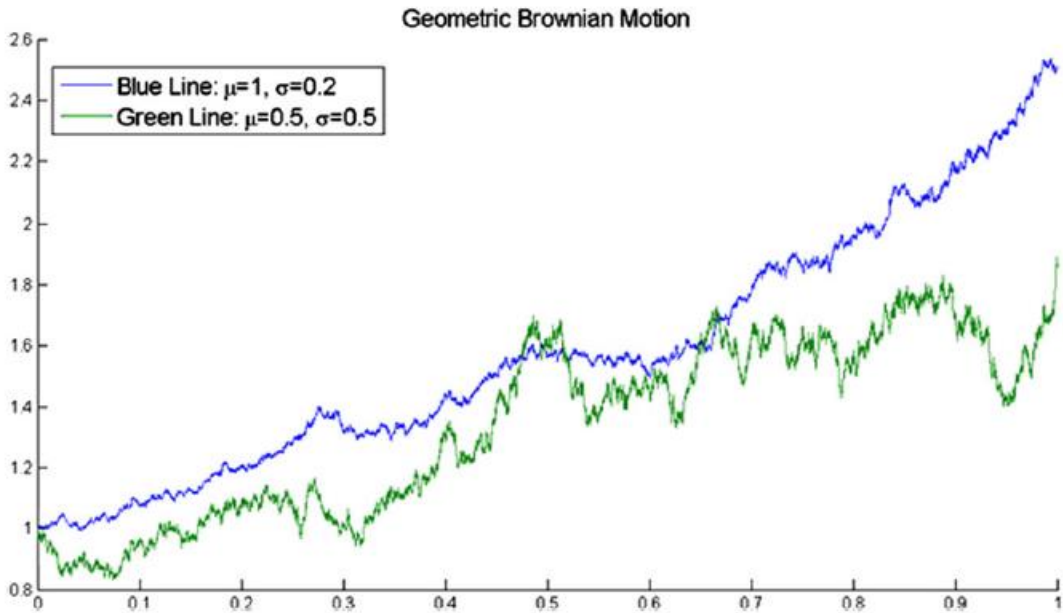

**Figure 2.** Two particular realizations of the stochastic process called **Geometric Brownian Motion (GBM)** taken from the Wikipedia site http://en.wikipedia.org/wiki/Geometric_Brownian_motion . Their mean values correspond to the exponential shown in red in Figure 1 (with other numeric values, for simplicity).

So, mathematicians assume that the *mean value* of the GBM is just the exponential red curve in Figure 1. That "marries" the exponentially increasing number of Living Species on Earth over 3.5 billion years with the need to take into account the *mass extinctions* that we know did take place in the past: like the "two main ones" occurred about 250 million years ago (Primary to Secondary Era transition) and 65 million years ago (dinosaurs' demise and transition between Secondary and Tertiary Eras, i.e. reptiles to mammals transition), plus "smaller" extinctions.

6) Please stop a moment to think about what our OVERCOME Theorem (Peak-Locus Theorem) means.

7) *Each new Species OVERCOMES the older Species.* Meaning that each new Species is a little more complex than the older Species from which it derives, but each new Species also is *more apt* than the older Species to face the difficulties of being alive. That is what Darwin called "natural selection", and that's the way we have cast Darwin's ideas into maths! Please note that we don't care about the specific DNA changes enabling the transition from an older Species to a newer one: that is the task of molecular biologists and related scientists. We just care about the overall mathematical picture of Evolution of Life on Earth over the last 3.5 billion years since… we later want to extrapolate that into the future and find out if Extraterrestrial Civilizations are living around us somewhere in the universe. That is SETI, the Search for ExtraTerrestrial Intelligence https://en.wikipedia.org/wiki/Search_for_extraterrestrial_intelligence.

8) So, what type of physical units shall we adopt for what is on the vertical axis in Figures 1 and 2 ? Just "the number of Living Species (as we just said) or something else "more profound" ? Well, we might think of *complexity* since each newer Species is more complex than the previous ones. Moreover, complexity theory has been studied by scholars over the past 50 years, and so we might refer to this book as "a book about the mathematics of complexity". But this author is primarily a mathematical physicist, and so he prefers to resort to the *information* unit that even children understand nowadays: bits, bytes, megabytes, gigabytes, terabytes and even petabytes, just like 1 petaByte is the number of raw data that SETI scientists had just now collected thanks to the "Breakthrough Listen" Project, to which this author is affiliated https://en.wikipedia.org/wiki/Breakthrough_Listen . As you have understood already, we want to see *information measured in bits* on the vertical axis in Figures 1 and 2.

9) *Information Theory* is the wonderful achievement of Claude Elwood Shannon (1916-2001), please see https://en.wikipedia.org/wiki/Claude_Shannon. In particular, Shannon introduced in 1948 the notion of *entropy of a probability density*, that we will largely use in this book: we simply call it "Shannon entropy". Then, we discovered that *the Shannon entropy is the natural measure of the Evolution of Life on Earth, and on Exoplanets.* But… Shannon entropy… of which probability densities? Of the b-lognormals appearing in Figure 1, of course. Fortunately, the Shannon entropy of a b-lognormal is given by a simple formula expressing it in terms of the two free b-lognormal parameters, mu and sigma, and so it becomes possible to compute the Shannon entropy of the GBM having as mean value the red solid exponential in Figure 1.

10) **Evo-Entropy**. By Evo-Entropy we mean the Shannon entropy of the Evolution of Life on Earth (and on Exoplanets). Then, little by little in the years between 2010 and 2015, this author came to make an unexpected discovery: if one assumes that the growth of the number of species is exponential (as in Figure 1), then the corresponding Evo-Entropy is just **LINEAR !!!** In other words, *the Evo-Entropy of life forms grew just like a STRAIGHT LINE, ranging from zero at just 3.5 billion years ago, to 25.575 bits nowadays.*

11) The following Figure 3 shows the time Evolution of the Entropy of life forms **(Evo-Entropy)** as a straight line ranging between 0 at the time of the origin of Life on Earth (3.5 billion years ago) and *25.575 bit nowadays (zero time).* We will later discover that this straight line is not exactly the Shannon entropy: actually we had to drop the minus sign in front of the traditional Shannon entropy definition (someone calls **"negentropy"** this "signed-reversed Shannon entropy) and we had to add a constant (i.e. time-independent term) to the traditional Shannon entropy definition in order to get the neat straight line growing from zero to 25.575 bits as shown in Figure 3. But these are "marginal details" good for nerds.

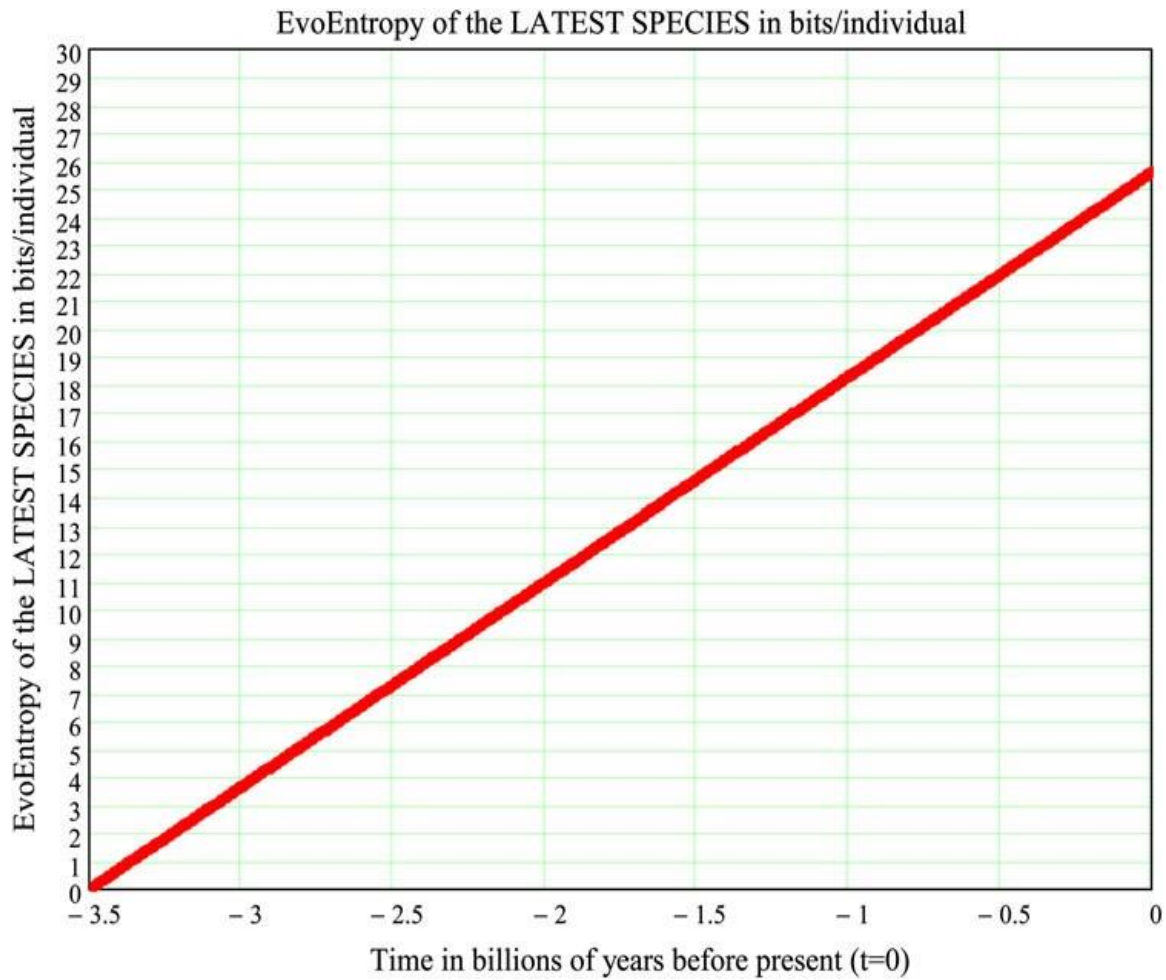

**Figure 3. Evo-Entropy** (in bits per individual) of the latest Species appeared on Earth over the last 3.5 billion years. This shows that a Man (nowadays) is 25.575 bits more evolved than was the first form of life, i.e. "something capable of reproducing itself" (that is RNA, in our assumption) about 3.5 billion years ago (it might be 3.8, but worry not!).

12) So, the above Figure 3 shows the time Evolution of the Entropy of life forms **(Evo-Entropy)** as a straight line ranging between 0 at the time of the origin of Life on Earth (3.5 billion years ago) and ***25.575 bit nowadays (zero time).*** This innocent-looking Figure 3, however, hides a much more profound meaning that we now wish to point out: Figure 3 is our "mathematical proof" that the Molecular Clock indeed is a fundamental law of nature. All biologists nowadays know about the Molecular Clock https://en.wikipedia.org/wiki/Molecular_clock, discovered in 1962 by Émile Zuckerkandl (1922-2013) and Linus Pauling (1901-1994). Their discovery paved the way to the Neutral Theory of Molecular Evolution (https://en.wikipedia.org/wiki/Neutral_theory_of_molecular_evolution) by Moto-o Kimura (1924-1994) and others, basically saying that "molecular evolution obeys the laws of quantum physics and not the natural selection of Darwin". Thus, molecular evolution on Earth must be the same on exoplanets too: not a small feat at all, confirming the idea that life must be present "everywhere in the universe".

13) And so we now reach the most profound consequence of the straight line in Figure 3. This is our **Evo-SETI SCALE for evolution of life in the universe.** The unit of the scale is called EE (Earth Evolution) and equals 25.575 bits. For a planet like Mars, the value on this scale is much smaller than 1 EE (if life actually appeared on Mars) and zero if no life ever existed on Mars. But for an exoplanet hosting a civilization more advanced than Human's on Earth nowadays, the value of EvoEntropy must be higher or much higher than 1 EE.

## 2. NASA vs. SETI: 1992-2020

We have a picture for you (this author is the second person to the left of the screen, "toasting" with his coffee cup).

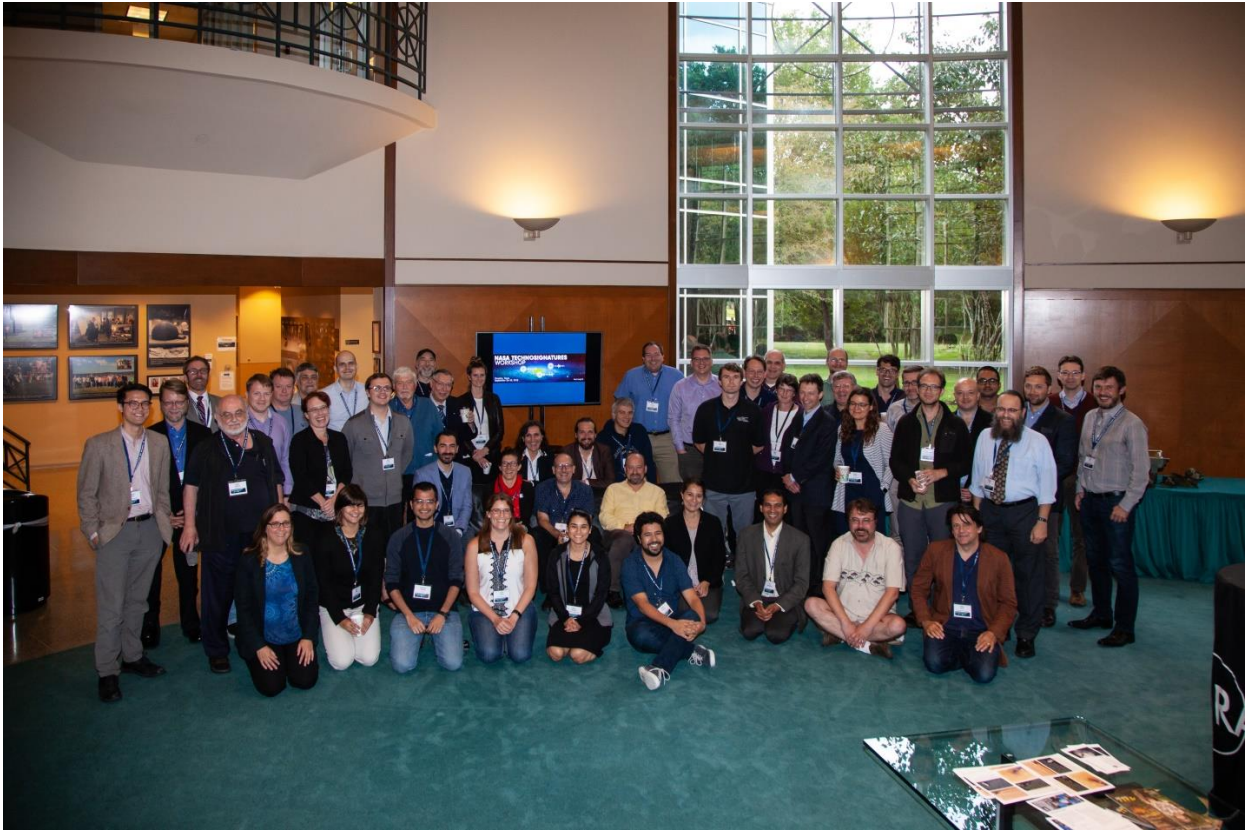

**Figure 4**. This picture was shot on September 27th, 2018, at the Lunar and Planetary Institute (LPI) in Houston, Texas, and shows the about 50 participants invited by NASA to attend the NASA Technosignatures Workshop described in detail at the web site https://www.hou.usra.edu/meetings/technosignatures2018/. The word "Technosignatures" seems rather cumbersome, but it really means SETI, the Search for ExtraTerrestrial Intelligence, advocated since 1959 by great minds like Giuseppe Cocconi, Phil Morrison, Frank Drake, Carl Sagan, Jill Tarter, Nikolay Kardashev and many more.

This is not the first time NASA takes an interest into SETI. The NASA SETI Program actually started on October 12th, 1992, exactly 500 years after Christopher Columbus landed in America for the first time. The person writing these lines was then aged 44 and was already deeply involved with SETI: he had the privilege to embark on one of the twelve buses that left Pasadena, California (the town where Caltech and NASA-JPL are located) on October 11th, and get to Barstow by the night. There, Carl Sagan gave the dinner speech. On the following morning we reached the site of NASA's Deep Space Network antenna at Goldstone, in the Mojave Desert. This antenna is shown in Figure 4 of the JBIS 1999 paper written by Steven J. Garber of the NASA History office and now available at the site https://history.nasa.gov/garber.pdf. There Carl Sagan officially opened NASA's All-Sky-Survey SETI search, while, at the same time, at the Arecibo radio telescope in Puerto Rico (at the time the world's largest) Bernard "Barney" Oliver, John Billingham, John Rummel, Jill Tarter, Seth Shostak and other SETI professionals officially opened the NASA Targeted Search, a SETI search on 778 stars of solar type nearby the Sun in the Galaxy.

Good old days.

But in less than a year's time the NASA SETI Program was over, terminated by Congress, as per the description given by Steven J. Garber in the JBIS 1999 paper mentioned above.

What happened next? Well, it happened that "people of good will" kept SETI alive in spite of all difficulties. And not just in the USA only. In Russia, France, Italy, Australia and Argentina, were a few "heroic radio astronomers" willing to "risk their good reputation" by doing SETI searches rather than just ordinary astrophysical work. This situation lasted roughly until 2010, when Britain's Astronomer Royal, Lord Martin Rees, declared that "absence of evidence is not evidence of absence". In the meantime, the discovery of more and more planets orbiting around other stars (exoplanets), and the success of space missions looking for such exoplanets, like NASA's "Kepler", convinced the various scientific establishments of leading countries that SETI was "respectable".

But in 2015 a new SETI revolution occurred: Russian-born American billionaire Yuri Milner gathered a meeting at the Royal Society in London and declared that he was willing to pay $100 million in the next 10 years for SETI Searches to be conducted at an unprecedented scale by his Breakthrough Listen (BL) Program (https://breakthroughinitiatives.org/initiative/1 ) taking place at the Department of Astronomy of the University of California at Berkeley (U. C. Berkeley).

Then, an explosion of SETI searches started. Several American Universities followed U. C. Berkeley's example, and so did some other countries, like Italy, where this author lives, and the U.K., France, the Netherlands, Russia, and especially China, that, in 2017, opened up her brand-new, 500-meter large FAST radio telescope https://drive.google.com/file/d/0B3YGjLNLQSmtZEowOE1ISHAxejg/view , now the world's largest.

So much for radio SETI, traditionally based on the assumption that Aliens would transmit radio messages around the hydrogen line of having 21 centimeters in wavelength, or, equivalently, 1420 MHz in frequency https://en.wikipedia.org/wiki/Hydrogen_line . But what if Aliens transmit at other frequencies? Well, if they send laser beams around, then Optical SETI might possibly detect these, as per the popular description at the site https://www.centauri-dreams.org/2017/07/17/detection-possibilities-for-optical-seti/ .

So much for ongoing SETI searches worldwide as of 2020.

### 3. Physics, Chemistry and Mathematics are THE SAME all over the Universe

Is Humankind ready for such a Contact with Aliens ?
Most probably not so yet, in this author's view.

We actually know "nothing" about them. We don't know how far they live from us. Most important, we don't know how much more technologically advanced than us they might possibly be. And that's scary.

Knowing nothing about Aliens, this author tried to resort to mathematics. This is because "two and two makes four everywhere in the universe", i.e. for both us and ExtraTerrestrials (abbreviated ETs, in the sequel). In fact, we know for sure that fundamental physics and chemistry are the same all over the universe, and we know so because the spectral lines in the light reaching us from the stars are the same as ours. But there is more. The frequencies of these spectral lines are computed on Earth by virtue of quantum mechanics. Thus, also quantum mechanics must be the same for both us and Aliens. In turn, quantum mechanics is part of high-level mathematics (eigenvalues, eigenvectors and the Hilbert space https://en.wikipedia.org/wiki/Hilbert_space ) and so the equations we scientists work with every day with must be the same for both us and Aliens. In other words still, not only physics and chemistry are the same for both us and ETs, but mathematics too has got to be the same. And some "philosophers" denying so are just (sorry!) too ignorant about modern science, and we will not pay attention to them any longer.

So, let's use mathematics, the only product of the human mind that may enable us to make further discoveries even before the relevant experimental counterpart is found. Our motto is "To measure is to understand" https://www.goodreads.com/quotes/632992-measurement-is-the-first-step-that-leads-to-control-and .

## 4. CLASSICAL Drake equation (1961) vs. STATISTICAL Drake equation (2008)

The foundational equation of SETI is the Drake equation (1961) https://en.wikipedia.org/wiki/Drake_equation. It was generalized by this author into the statistical Drake equation in 2008, as described in Chapter 1 of this book as well as at the Wikipedia site https://link.springer.com/chapter/10.1007/978-3-642-27437-4_1.
Both the classical equation of Frank Drake (1961) and the Statistical equation of Claudio Maccone (2008) estimate the number *N* of communicating civilizations now existing in our galaxy, the Milky Way, but they do so in different ways: for Drake, *N* is the product of the seven positive numbers listed at the Drake equation site given earlier. For Maccone, *N* is the is the product of a very large number of factors (theoretically speaking, an "infinite" number of factors), each of which is a probability density function (abbreviated "pdf") having its own mean value and its own standard deviation, both known to scientists by virtue of experimental measurements. Then, whatever the probability density functions might possibly be, the so-called Central Limit Theorem of Statistics shows that the probability density function of *N* is a **lognormal**. More about lognormals and b-lognormals in the next section.

## 5. The b-LOGNORMAL EQUATION (i.e. probability density function = pdf), easily proved as follows

This book is based on the fundamental notion of a lognormal pdf (in the time), with the consequent *b*-lognormal pdf in the time, that is just a lognormal starting at the initial instant *b*=birth other than zero.

To let even the Preface to this book be self-contained in this regard, we now provide an easy proof of the *b*-lognormal equation as a probability density function (pdf). Just start from the well-known Gaussian or normal pdf

$$\text{Gaussian\_or\_normal\_pdf}(x;\mu,\sigma) = \frac{e^{-\frac{(x-\mu)^2}{2\sigma^2}}}{\sqrt{2\pi}\,\sigma} \quad \text{with} \quad \begin{cases} -\infty < x < +\infty, \text{ independent variable.} \\ -\infty < \mu < +\infty, \text{ a real parameter} = \text{where the peak is.} \\ \sigma > 0, \text{ a positive parameter} = \text{the standard deviation.} \end{cases} \quad (1)$$

This pdf has two parameters:
1) $\mu$ turns out to be the mean value of the Gaussian and the abscissa of its peak. Since the independent variable $x$ may take up any value between $-\infty$ and $+\infty$, i.e. it is a real variable, so $\mu$ must be real too.
2) $\sigma$ turns out to be the standard deviation of the Gaussian and so it must be a positive variable.
3) Since the Gaussian is a pdf, it must fulfill the normalization condition

$$\int_{-\infty}^{\infty} \frac{e^{-\frac{(x-\mu)^2}{2\sigma^2}}}{\sqrt{2\pi}\sigma} dx = 1 \quad (2)$$

and this is the equation we need in order to "discover" the *b*-lognormal. Just perform in the integral (2) the substitution $x = \ln(t)$ (where ln is the natural logarithm, i.e. the one to base e = 2.718281828459045… ). Then (2) is turned into the new integral

$$\int_0^\infty \frac{e^{-\frac{(\ln(t)-\mu)^2}{2\sigma^2}}}{\sqrt{2\pi}\,\sigma\,t} dt = 1. \quad (3)$$

But this (3) may be regarded as the normalization condition of another random variable, ranging "just" between zero and $+\infty$, and this new random variable we call "lognormal" since it "looks like" a normal one except that $x$ is now replaced by $x = \ln(t)$ and *t* now also appears at the denominator of the fraction. In other words, the lognormal pdf is

$$\text{lognormal\_pdf}(t; \mu, \sigma) = \frac{e^{-\frac{(\ln(t) - \mu)^2}{2\sigma^2}}}{\sqrt{2\pi} \cdot \sigma \cdot t} \quad \text{with} \quad \begin{cases} 0 \leq t < +\infty, \text{ independent variable.} \\ -\infty < \mu < +\infty, \text{ a real parameter (no special name).} \\ \sigma > 0, \text{ a positive parameter (no special name).} \end{cases} \quad (4)$$

At the site https://en.wikipedia.org/wiki/Log-normal_distribution various plots of (4) are shown.

Just one more step is required to jump from the "ordinary lognormal" (4) (i.e. the lognormal starting at *t* = 0) to the *b*-lognormal, that is the lognormal starting at any real instant *b* ("*b*" stands for "birth"). Since this simply is a shifting along the time axis from 0 to the new real time origin *b*, in mathematical terms it means that we have to replace *t* by (*t* – *b*) everywhere in the pdf (4). Thus the *b*-lognormal pdf must have the equation

$$\text{b-lognormal\_pdf}(t; \mu, \sigma, b) = \frac{e^{-\frac{(\ln(t-b) - \mu)^2}{2\sigma^2}}}{\sqrt{2\pi} \cdot \sigma \cdot (t-b)} \quad \text{with} \quad \begin{cases} b < t < +\infty, \text{ time = real independent variable starting at } b. \\ -\infty < \mu < +\infty, \text{ a real parameter (no special name).} \\ \sigma > 0, \text{ a positive parameter (no special name).} \\ b = \text{birth time, the (real) time when the b-lognormal starts.} \end{cases} \quad (5)$$

Sometime, the *b*-lognormal (5) is called "three-parameter lognormal" by statisticians. That is correct since the three parameters appearing in (5) are $(\mu, \sigma, b)$. However, we prefer to call it *b*-lognormal to stress its biological meaning as the probability density representing the LIFETIME of any living being, born at the instant *b*.

We will later use *b*-lognormals also to represent the lifetime of Historic Civilizations too, like the nine historic Civilizations that will be studied in the sequel of this Preface.

### 6. Assume LIFE on Earth STARTED 3.5 BILLION YEARS AGO: then *ts* = -3.5 10$^9$ years, and zero time is now

Question: is it possible to measure mathematically the Biological Evolution (also called Darwinian Evolution) that paleontologists tell us occurred on Earth over (about) the last 3.5 billion years? Our answer is "yes" as we show that in this book.

Darwin (1859, "The Origin of Species") had the great merit to dispel all religious assumptions from the biological sciences by realizing that animals and plants evolved from primitive forms up to humans over that long amount of time. How long exactly? Radioactivity gave the relevant answer starting about 1910, after Darwin had passed away in 1882. Here again mathematics plays a key role in radioactivity, as given by the Harry Bateman's analytical solution to the experimental differential equations discovered by physicists Pierre and Marie Curie and others, see https://en.wikipedia.org/wiki/Bateman_equation. So, the "long time" of Evolution that Darwin did not know about was estimated over the last 100 years by paleontologists, biologists, geneticists and other scientists to be over 3.5 billion years, possibly 3.8, or even 4 (as someone claims), given that the Earth originated from the Sun about 4.5 billion years ago (like all planets in the Solar System).

In all chapters of this book, we shall assume for simplicity that life on Earth started exactly 3.5 billion years ago, just to fix the ideas. This number will be denoted by *ts* (= time of start [of life on Earth]) in all equations. So, *ts* = -3.5 10$^9$ years, assuming that the zero-time is nowadays, i.e. that times prior to nowadays are denoted by negative numbers (in years), while future times will be positive numbers for us.

The first form of life, biologists say nowadays, was RNA, and DNA then followed. We all know that DNA is "the molecule of life", meaning that every baby is born out ofthe merging of the father's and mother's DNA. But DNA is the same for all living members of a certain living Species. Darwin could hardly understand what's happening when a new Species is originated by an older Species: the answer was found after 1900 by the British school of mathematical geneticists (Francis Galton, Karl Pearson, Godfrey Hardy, Ronald Fisher, J. B. S. Haldane and Francis Crick, just to name a few). Many American geneticists also contributed after about 1900.

In this book we assume that the number of living Species on Earth over the last 3.5 billion years is well represented, mathematically speaking, by a Geomeric Brownian Motion (GBM). In fact:

1) We know that the first living Species (meaning the first molecule capable of *reproducing* itself) was RNA 3.5 billion years ago. In the language of mathematics, this is the same as saying that the initial condition of our GBM is one, that is *Ns*=1 at *ts*=-3.5*10^9 years.
2) Nowadays, i.e. at the time *t*=0 in our time convention, the supposed number of Species living on Earth is about 50 million (so many biologists guess, including insect Species in the count) but this number could be quite higher if we include Bacteria in the count. At the moment we shall assume *Ne*=50 million, where *Ne* means the number of living Species at the *end* of the observed timespan, i.e. nowadays.
3) Then the Geometric Brownian Motion (GBM, site https://en.wikipedia.org/wiki/Geometric_Brownian_motion ) is our way to cast Biological Evolution mathematically, i.e. by virtue of just a few, simple statistical equations.
4) Figures 1 and 2 summarize all that in just two plots!
5) In the English-speaking world, this plots also goes under the name of "Malthusian growth", since Thomas Malthus (1766-1834) used a similar exponential curve in 1798 to describe the human population growth (https://en.wikipedia.org/wiki/Malthusian_growth_model). But we prefer to use the term "exponential growth".

In this book we assume that the number of living Species on Earth over the last 3.5 billion years is well represented, mathematically speaking, by a GBM.

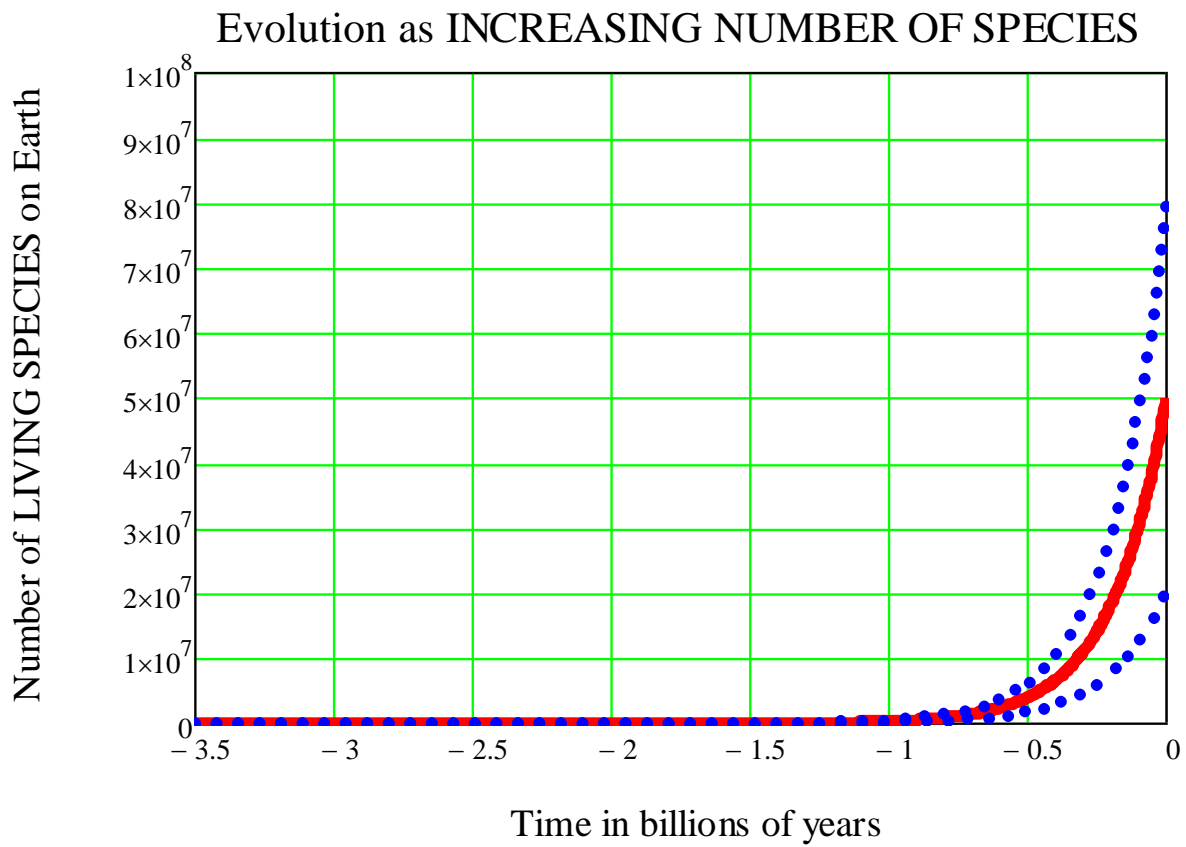

**Figure 5.** BIOLOGICAL EVOLUTION as the increasing number of living Species on Earth between 3.5 billion years ago and now. The red solid curve is the exponential mean value $m_{\text{GBM}}(t)$ of the Geometric Brownian Motion (GBM) stochastic process $L_{\text{GBM}}(t)$, while the blue dot-dot curves above and below the exponential mean value are the two standard deviation upper and lower curves. The "Cambrian Explosion" of life, that on Earth started around 542 million years ago, is evident in the above plot just before the value of -0.5 billion years in time, where all three curves "start leaving the time axis and climbing up". Notice also that the starting value of living species 3.5 billion years ago

is ONE by definition, but it "looks like" zero in this plot since the vertical scale (which is the true scale here, not a log scale) does not show it. Notice finally that nowadays (i.e. at time $t=0$) the two standard deviation curves have exactly the same distance from the middle mean value curve, i.e. 30 million living species more or less the mean value of 50 million species. These are assumed values that we used just to exemplify the GBM mathematics: biologists might assume other numeric values. But our describing equations are going to remain the same.

### 7. How to take EXTINCT SPECIES into account by virtue of the stochastic process called (incorrectly) "GEOMETRIC BROWNIAN MOTION" (GBM)

Mathematically naïve folks might object that many Species went extinct during the 3.5 billion years of Biological Evolution, and so representing Evolution by virtue of a simple, increasing exponential in the time is an oversimplified view. Well, this objection is easily answered by mathematicians: Biological Evolution is not exactly the exponential shown above in red, but is rather a stochastic process (i.e. a random function of the time) the mean value of which is indeed the above exponential.

Wall Street economists have studied that stochastic process since about fifty years ago to represent mathematically the fair price or theoretical value for a call or a put option based on six variables such as volatility, type of option, underlying stock price, time, strike price, and risk-free rate. These mathematical studies are called Black-Scholes models (https://en.wikipedia.org/wiki/Black%E2%80%93Scholes_model ) and the Nobel prize in Economics was awarded in 1997 to Merton and Scholes (William Black had unfortunately passed away already).

So, what this author did, in the practice, was to pick up the GMB used in the Black-Scholes model and re-use it to describe Biological Evolution over the last 3.5 billion years.

With a caveat: this GBM is not what physicists call a Brownian motion, so tune up your words according to the person you are talking to: in physics, a Brownian motion is a Gaussian stochastic process, while a Wall Street's GBM actually is a lognormal stochastic process, i.e. a process of the type e =2.71828… raised to the Gaussian process. Hard to explain by words, but easy by equations, as we shall see in this book.

### 8. Our earlier mathematical DEFINITION OF "LIFETIME" of a LIVING BEING: IT IS A "b-LOGNORMAL" IN THE TIME

The first significant innovation put forward by this author in his Evo-SETI Theory was his definition of lifetime of a living being as a b-lognormal in the time. As per equation (5), a b-lognormal is a lognormal in between the starting point (*b* = birth) and the "senility" point *s*=the time of the lognormal's descending inflexion, as shown in Figure 6. After the instant *s* the curve is a descending straight line until it reaches the time axis again at the "death time" *d*.

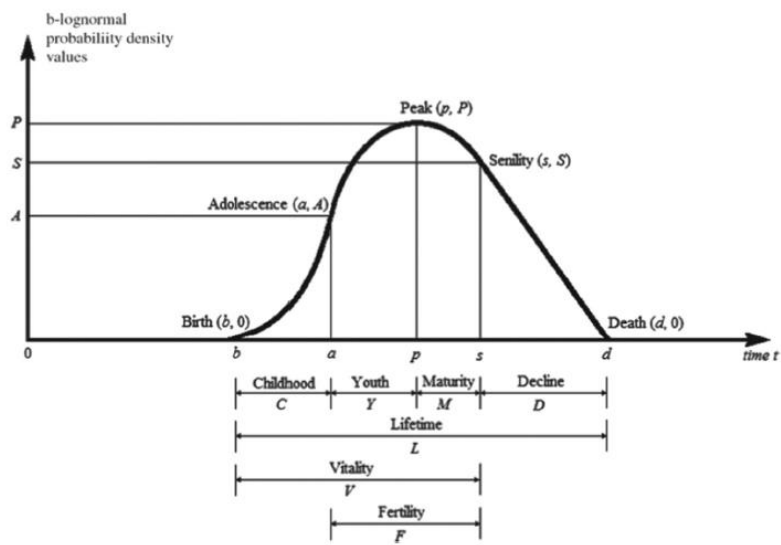

**Figure 6.** Lifetime of every living being as a b-lognormal between birth *b* and death *d*, with a junction at senility *s*. More definitions like Childhood, Youth, Maturity, Decline, Vitality and Fertility are just obvious consequences.

## 9. Our first innovative discovery made in 2010: the TWO b-LOGNORMAL HISTORY FORMULAE

The most important mathematical consequences of our definition of life are the following two b-LOGNORMAL HISTORY FORMULAE expressing the two parameters $\sigma$ and $\mu$ of the b-lognormal pdf in terms of the three parameters $(b, s, d)$ of birth, senility and death, respectively. They were the first important step ahead in Evo-SETI Theory made by this author, and were discovered by him at a date prior to March 10, 2011 (the full proof is given at the pages 172-181 in Chapter 6, Appendix 6.B, of this author's 2012 book "Mathematical SETI"):

$$\begin{cases} \sigma = \dfrac{d-s}{\sqrt{d-b} \cdot \sqrt{s-b}} \\ \mu = \ln(s-b) + \dfrac{(d-s)(d+b-2s)}{(d-b)(s-b)} \end{cases} \quad (6)$$

In other words, if one assigns the birth, senility and death time of a living being, then the corresponding b-lognormal plot is found immediately. The proof is so easy that even a good freshman should be able to find it!

## 10. History of 9 "Western" Historic Civilizations as b-LOGNORMALS: Ancient Egypt to USA

The unexpected bonanza of the History Formulae (6) is that they apply not only to the lifetime of any living being: they also apply to the *lifetime of Historic human Civilizations.* Just look at the following Figure 7, please.

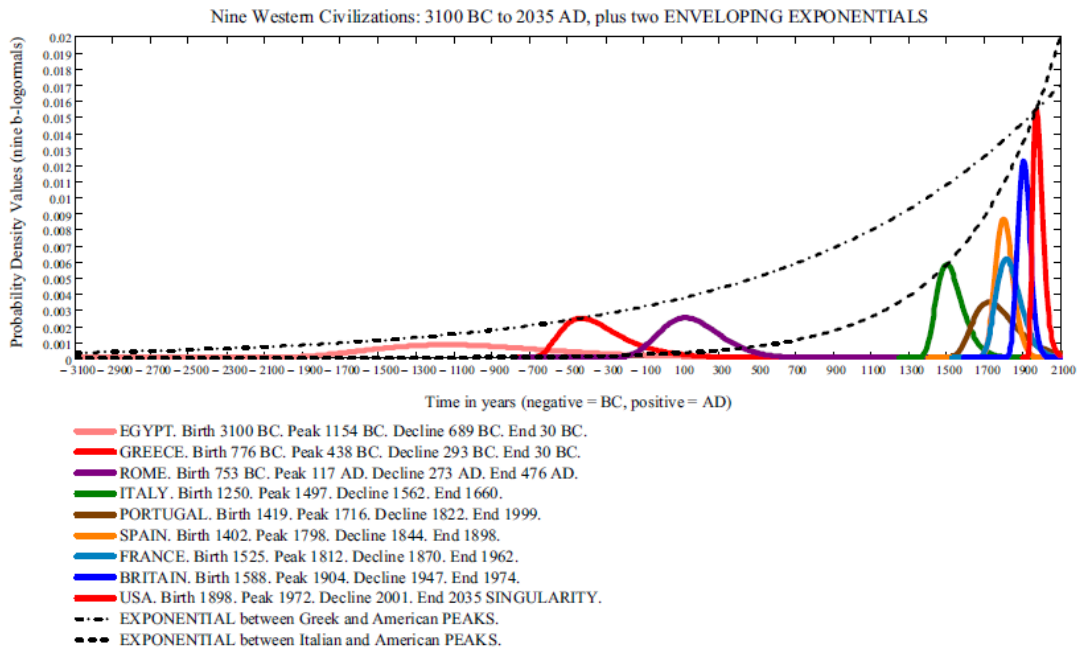

108          C. Maccone / Acta Astronautica 122 (2016) 106–113

Nine Western Civilizations: 3100 BC to 2035 AD, plus two ENVELOPING EXPONENTIALS

— EGYPT. Birth 3100 BC. Peak 1154 BC. Decline 689 BC. End 30 BC.
— GREECE. Birth 776 BC. Peak 438 BC. Decline 293 BC. End 30 BC.
— ROME. Birth 753 BC. Peak 117 AD. Decline 273 AD. End 476 AD.
— ITALY. Birth 1250. Peak 1497. Decline 1562. End 1660.
— PORTUGAL. Birth 1419. Peak 1716. Decline 1822. End 1999.
— SPAIN. Birth 1402. Peak 1798. Decline 1844. End 1898.
— FRANCE. Birth 1525. Peak 1812. Decline 1870. End 1962.
— BRITAIN. Birth 1588. Peak 1904. Decline 1947. End 1974.
— USA. Birth 1898. Peak 1972. Decline 2001. End 2035 SINGULARITY.
-·-· EXPONENTIAL between Greek and American PEAKS.
··· EXPONENTIAL between Italian and American PEAKS.

**Figure 7.** Lifetime of 9 Western Civilizations as b-lognormals. From Ancient Egypt, Greece and Rome, to the Italian Renaissance and then the colonial empires of Portugal, Spain, France, Britain and the USA, the History Formulae (6) are just all that is required to represent History mathematically: you just give the numbers for *b*, *s*, and *d* for each Civilization, as given in Table 1, and the 9 above b-lognormal plots come out immediately! Note that, along the vertical axis, are just real positive numbers and nothing else. In fact, the b-lognormals are just probability densities, i.e. curves staying only above the time axis. As easy as that!

So, let us review our History inputs by the following Table 1. There are $3 \cdot 9 = 27$ input numbers (27 real numbers).

**Table 1**
Birth, peak, decline and death times of nine Historic Western Civilizations (3100 BC–2035 AD), plus the relevant peak heights. They are shown in Fig. 2 as nine b-lognormal probability density functions (pdfs).

| | $b$=Birth time | $p$=Peak time | $s$=Decline = senility time | $d$=Death time | $P$=Peak ordinate |
|---|---|---|---|---|---|
| Ancient Egypt | 3100 BC<br>Lower and Upper Egypt unified.<br>First Dynasty. | 1154 BC<br>Luxor and Karnak temples edified by Ramses II by 1260 BC. | 689 BC<br>Assyrians invade Egypt in 671 BC, leave 669 BC. | 30 BC<br>Cleopatra's death: last Hellenistic queen. | $8.313 \cdot 10^{-4}$ |
| Ancient Greece | 776 BC<br>First Olympic Games, from which Greeks compute years. | 434 BC<br>Pericles' Age.<br>Democracy peak.<br>Arts and Science peak.<br>Aristotle. | 323 BC<br>Alexander the Great dies.<br>Hellenism starts in Near East. | 30 BC<br>Cleopatra's death: last Hellenistic queen. | $2.488 \cdot 10^{-3}$ |
| Ancient Rome | 753 BC<br>Rome founded.<br>Italy seized by Romans by 270 BC, Carthage and Greece by 146 BC, Egypt by 30 BC. Christ 0. | 117 AD<br>Rome at peak: Trajan in Mesopotamia.<br>Christianity preached in Rome by Saints Peter, Paul against slavery by 69 AD. | 273 AD<br>Aurelian builds new walls around Rome after Military Anarchy, 235–270 AD. | 476 AD<br>Western Roman Empire ends.<br>Dark Ages start in West.<br>Not in East. | $2.193 \cdot 10^{-3}$ |
| Renaissance Italy | 1250<br>Frederick II dies.<br>Middle Ages end. Free Italian towns start Renaissance. | 1497<br>Renaissance art and architecture. Birth of Science. Copernican revolution (1543). | 1564<br>Council of Trent ends.<br>Catholic and Spanish rule. | 1660<br>Cimento shut. Bruno burned 1600. Galileo died 1642. | $5.749 \cdot 10^{-3}$ |
| Portuguese Empire | 1419<br>Madeira island Discovered. African coastline explored by 1498. | 1716<br>Black slave trade to Brazil at its peak. Millions of blacks enslaved or killed. | 1822<br>Brazil independent, other colonies retained. | 1999<br>Last colony, Macau, lost to Republic of China. | $3.431 \cdot 10^{-3}$ |
| Spanish Empire | 1402<br>Canary islands are conquered by 1496. In 1492 Columbus discovers America. | 1798<br>Largest extent of Spanish colonies in America: California settled since 1769. | 1805<br>Spanish fleet lost at Trafalgar. | 1898<br>Last colonies lost to the USA. | $5.938 \cdot 10^{-3}$ |
| French Empire | 1524<br>Verrazano first in New York bay.<br>Cartier in Canada, 1534. | 1812<br>Napoleon I dominates continental Europe and reaches Moscow. | 1870<br>Napoleon III defeated.<br>Third Republic starts.<br>World Wars. | 1962<br>Algeria lost as most colonies.<br>Fifth Republic starts in 1958. | $4.279 \cdot 10^{-3}$ |
| British Empire | 1588<br>Spanish Armada Defeated.<br>British Empire's expansion starts. | 1904<br>British Empire peak. Top British Science: Faraday, Maxwell, Darwin, Rutherford. | 1947<br>After World Wars One and Two, India gets independent. | 1974<br>Britain joins the EEC and loses most of her colonies. | $8.447 \cdot 10^{-3}$ |
| USA Empire | 1898<br>Philippines, Cuba, Puerto Rico seized from Spain. | 1972<br>Moon Landings, 1969–72: America leads the world. | 2001<br>9/11 terrorist attacks: decline. Obama 2009. | 2035<br>Singularity?<br>Will the USA yield to China? | 0.013 |

**Table 1**. You just input the three numbers *b*, *s* and *d* for each Civilization (i.e. b-lognormal), and the 9 plots shown in Figure 7 come out automatically. Isn't this surprising? Well, even more surprising it would be if the SETI scientists would discover an Alien Civilization in space, and then we could figure out mathematically how much "more advanced than us" they are. This is precisely the goal of this this mathematical book, i.e. the goal of our Evo-SETI Theory, an acronym standing for "Evolution and SETI".

## 11. MOLECULAR CLOCK predicted by our Evo-SETI Theory as the LINEAR EVO-ENTROPY of the GMBs

What is a clock ?

Mathematically speaking, a clock is a straight line in a diagram with time on the horizontal axis and whatever changes LINEARLY in time on the vertical axis, like sand in an hourglass. Then Figure 3 is a clock.

Let us now consider the "molecular clock", i.e. the greatest experimental discovery of molecular biology if not of all of astrobiology (site: https://en.wikipedia.org/wiki/Molecular_clock ). We report here just the same the description of the molecular clock provided by Wikipedia: The **molecular clock** is figurative term for a technique that uses the mutation rate of biomolecules to deduce the time in prehistory when two or more life forms diverged. The biomolecular data used for such calculations are usually nucleotide sequences for DNA or amino acid sequences for proteins. The benchmarks for determining the mutation rate are often fossil or archaeological dates. The molecular clock was first tested in 1962 on the hemoglobin protein variants of various animals, and is commonly used in molecular evolution to estimate times of speciation or radiation.

So what ?

So, this author thinks that his Evo-SETI Theory embodies the existence of the Molecular Clock over 3.5 billion years since the appearance of Life on Earth: the Molecular Clock is nothing but the EvoEntropy (Shannon Entropy with a reversed sign in front, and starting at zero 3.5 billion years ago) of the Geometric Brownian motion, i.e. of the exponentially increasing number of Species over the same 3.5 billion years. Wow ! That's a discovery !

Let us make an historical comparison with two great scientists of the past: Johannes Kepler (1571-1630) and Isaac Newton (1642-1527). Kepler discovered the laws of planetary motion by translating mathematically all the careful observations of Mars made by his teacher Tycho Brahe (1546-1601) into three simple equations: https://en.wikipedia.org/wiki/Kepler%27s_laws_of_planetary_motion. But Newton did better still: he proved mathematically that the three Kepler's Laws where just mathematical consequences of his single Law, the Law of Universal Gravitation: https://en.wikipedia.org/wiki/Newton%27s_law_of_universal_gravitation . In other words, Newton's established a new branch of science, called Celestial Mechanics up to about 1950 and Astrodynamics after the advent of space missions.

Similarly, we claim that our Evo-SETI Theory has mathematized the Molecular Clock, and will have consequences in the future of Astrobiology that we can hardly envisage nowadays: it is too early yet.

## 12. INFORMATION GAPS among different Civilizations as ENTROPY DIFFERENCES among them

One of the possibilities of Evo-SETI Theory is to describe mathematically how much a Civilization is more advanced than another one by computing the difference among the two relevant EvoEntropies. Please look at the following Table 3

**Table 3**
INFORMATION GAPS=(Shannon) ENTROPY DIFFERENCES in bits/individual among the nine Historic Western Civilizations (3100 BC–2035 AD) shown in Fig. 2 and Table 1.

| INFORMATION GAP in bits/individual | Egypt | Greece | Rome | Italy | Portugal | Spain | France | Britain | USA |
|---|---|---|---|---|---|---|---|---|---|
| Egypt | 0 | 1.467 | 1.622 | 2.795 | 2.008 | 3.425 | 4.624 | 4.335 | 3.864 |
| Greece | −1.467 | 0 | 0.154 | 1.328 | 0.541 | 1.958 | 3.157 | 2.868 | 2.397 |
| Rome | −1.622 | −0.154 | 0 | 1.173 | 0.386 | 1.804 | 3.002 | 2.713 | 2.242 |
| Italy | −2.795 | −1.328 | −1.173 | 0 | −0.7872 | 0.630 | 1.829 | 1.540 | 1.069 |
| Portugal | −2.008 | −0.541 | −0.386 | 0.7872 | 0 | 1.418 | 2.616 | 2.327 | 1.856 |
| Spain | −3.425 | −1.958 | −1.804 | −0.630 | −1.418 | 0 | 1.198 | 0.909 | 0.438 |
| France | −4.624 | −3.157 | −3.002 | −1.829 | −2.616 | −1.198 | 0 | −0.289 | −0.759 |
| Britain | −4.335 | −2.868 | −2.713 | −1.540 | −2.327 | −0.909 | 0.289 | 0 | −0.471 |
| USA | −3.864 | −2.397 | −2.242 | −1.069 | −1.856 | −0.438 | 0.759 | 0.471 | 0 |

This is an antisymmetric matrix (also called skew-symmetric matrix) expressing the EvoEntropy GAPS (in bits/individual) among the nine Historic Western Civilizations described in Figure 3 and Table 1.

We are now able to understand the #1 success of Evo-SETI Theory: quantifying the GAP among different Civilizations by virtue of JUST ONE NUMBER. For instance, using data similar to the above Table 2, we proved that the Evo-Entropy gap between the Aztecs and the Spaniards when they clashed in 1519 was about 3.85 bits/individual. In other words, back in 1519 each Spaniard was, on the average, 3.85 bits more knowledgeable than any Aztecs. That's why the Aztecs had a psychological breakdown. So, Evo-SETI Theory could even be used in Mathematical Psychology, making that research field more quantitative than it has been up to now.

### 13. Introducing ENERGY into Evo-SETI Theory

Evo-SETI Theory, as described up to now, dealt only with (the Shannon) Entropy, and not with ENERGY at all. And that was the situation prior to the year 2015. However, both Entropy and Energy are the two pillars of classical Thermodynamics, and so we can hardly expect to create a good model for the evolution of life on Earth and Exoplanets if we just use Entropy only: we must insert Energy to Evo-SETI Theory too. How can we do so ?

The answer to this question was found by this author on November 22$^{nd}$, 2015, when he discovered his new "Logpar History Formulae", as described in the next and following sections.

### 14. Our 2015 new mathematical definition of "LIFETIME of a Living Being": a "LOGPAR" POWER CURVE in the time

On November 22$^{nd}$, 2015, this author made one more mathematical discovery. He replaced the b-lognormal described so far by virtue of another curve that he had never considered earlier: the "logpar". The acronym logpar stands for "LOGnormal plus PARabola" and means a curve that is a **b-lognormal only between birth and peak time**, but actually just a parabola between peak time and death time.

Let us see this LOGPAR plot in the case of the Civilization of Ancient Rome (753 BC to 476 AD).

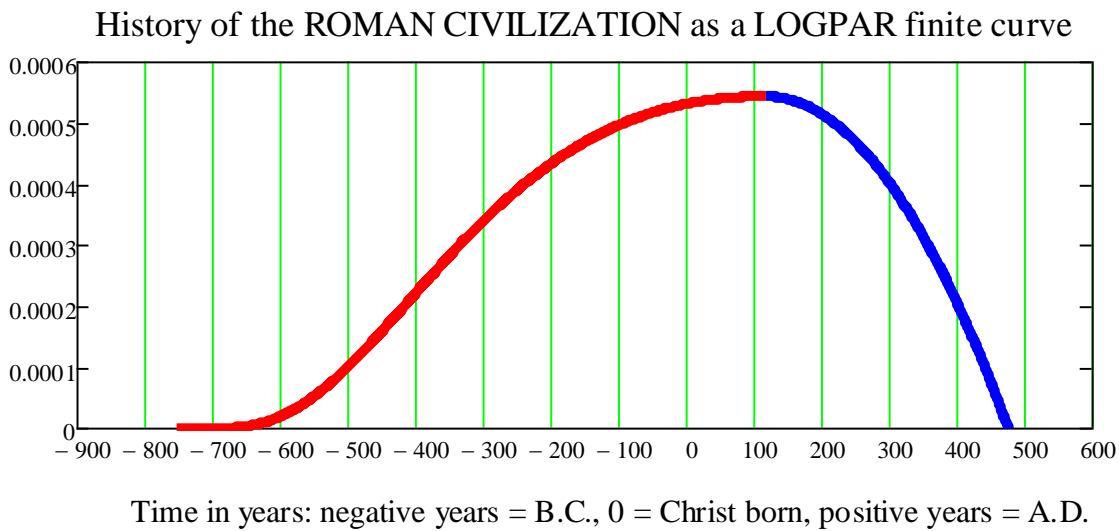

Time in years: negative years = B.C., 0 = Christ born, positive years = A.D.

**Figure 6.** Representation of the History of the Roman civilization as a LOGPAR finite curve: the ascent of Rome is shown in red (753 B.C. to 117 A.D.) and the decline in blue (117 A.D. to 476 A.D.). Rome was funded in 753 B.C., i.e. in the year -753 in our notation, or $b = -753$. Then the Roman republic and empire (the latter since the first emperor, Augustus, roughly after 27 B.C.) kept growing in conquered territory until it reached its peak (maximum extension, up to Susa in current Iran) in the year 117 A.D., i.e. $p = 117$, under emperor Trajan. Afterwards it started to decline and loose territory until the final collapse in 476 A.D. ($d = 476$, Romulus Augustulus, last emperor). Thus,

just three points in time are necessary to summarize the History of Rome: $b = -753, p = 117, d = 476$. No further intermediate point, like senility in between peak and death, is necessary since we now used a logpar rather than a b-lognormal (b-lognormals *only* had been used by this author in the years 2009-2017).

Evidently, the adoption of a logpar instead of a b-lognormal abolishes the senility instant *s* and replaces its role by the role of the peak time *p*. In the practice, it is much easier to estimate someone's peak time *p* than senility time *s*, and so abandoning b-lognormals in favour of logpars greatly simplifies things for the applications of Evo-SETI Theory to real cases. In other words, we will only have to specify the logpar triplet (*b, p, d*) in order to get an easily understandable logpar curve for all further calculations based on it.

### 15. Our two LOGPAR HISTORY FORMULAE

Thus, the real discovery that this author made on November 22$^{nd}$, 2015 was the discovery of the LOGPAR HISTORY FORMULAE, expressing the logpar's $\sigma$ and $\mu$ in terms of the "easy triplet" *b*, *p*, *d*, and **not requesting *s* any more** :

$$\begin{cases} \sigma = \dfrac{\sqrt{2}\sqrt{d-p}}{\sqrt{2d-(b+p)}} \\ \mu = \ln(p-b) + \sigma^2 \end{cases} \tag{7}$$

But that is just the most evident change. Other more subtle mathematical changes, that become evident only by doing the calculations, are:
1) We have **dropped** the hypothesis that life is a **probability density** as it was the case for the b-lognormals. In other words, the logpar curves are positive curves **not** fulfilling the normalization condition requesting that the area under the b-lognormal (i.e. the area under the curve in between *b* and $+\infty$) must equal one. In fact…
2) Logpars are now **positive power curves** measured in Watts. With this interpretation, the area under each logpar, i.e. the integral (with respect to the time, of course) of the logpar in between the birth *b* and any instant *z* before death *d*, is the ENERGY necessary to that living being to live up to the instant *z* (called "progressive energy").
3) Consequently, the LIFETIME ENERGY of that living being is the "final" progressive energy up to death time *d*.
4) A further key result provided by logpars (and *not* y b-lognormals) is the **"Principle of Minimum Energy"**, saying that the derivation of the logpar history formulae (7) is possible only under the assumption that Lifetime Energy described at point 3) is a **minimum** with respect to the positive parameter $\sigma$. In words, this mathematical discovery very much resembles the "principle of least action" of paramount importance in Physics. But we confess that we still have to dig more profoundly about this Principle in future mathematical papers. The potential consequences of this Minimum Energy Principle might be enormous for the applications of Evo-SETI Theory to the study of all Living Beings, but at the moment (January 2020) all this still is premature.

### 16. Approaching IMMORTALITY, that is letting the death-time *d* approach infinity

One of the classical themes of Science Fiction is "Living Beings Approaching Immortality". Actually, the life expectancy of humans has already increased considerably during the last few decades, so one may wonder whether, in the far future, humans will become "nearly immortals" or so. Evo-SETI Theory offers a mathematical clue about immortality: just let the death time approach infinity, that is let $d \to \infty$ in the Logpar History Formulae (7), and see what happens. Well, it happens that $\sigma \to 1$ and this is a striking novelty with respect to b-lognormals, where $0 < \sigma < \infty$ still in accordance with the meaning of $\sigma$ as standard deviation of the Gaussian = normal distribution. In other words, having abandoned the normalization condition in the transition from b-lognormals to logpars, now pays off for further improvements. For instance, what happens if we let $d \to \infty$ in the Ancient Rome logpar plot in Figure 6 ? The result is most surprising and it is shown in the following Figure 7, as we now describe.

## 17. LOGPAR ENERGY's approach to IMMORTALITY "in the long run"

Figure 7 shows not only "The Decline and Fall of the Roman Empire" (117-378 A.D., just to use Edward Gibbon's great book's title), but also the Dark Ages (Middle Ages), say, in between 476 A.D. and the Italian Renaissance (after roughly 1300 A.D.): it took 1000 years to the European Western Civilization to recover from the Barbaric Invasions within the Roman Empire firstly, and then within the Romance Countries secondly. The solid red line is Figure 7 is the Energy of the once Roman lands, normalized to the ½ value at the peak time energy of Trajan in 117 A.D.

But Figure 7 also shows the oblique asymptote (dot-dot straight line) that the above Energy approaches at higher and higher values of the time: the same Trajan normalized value of ½ was reached again during the Italian Renaissance around 1523 A.D. After that, Energy increased like a straight line (the asymptote).

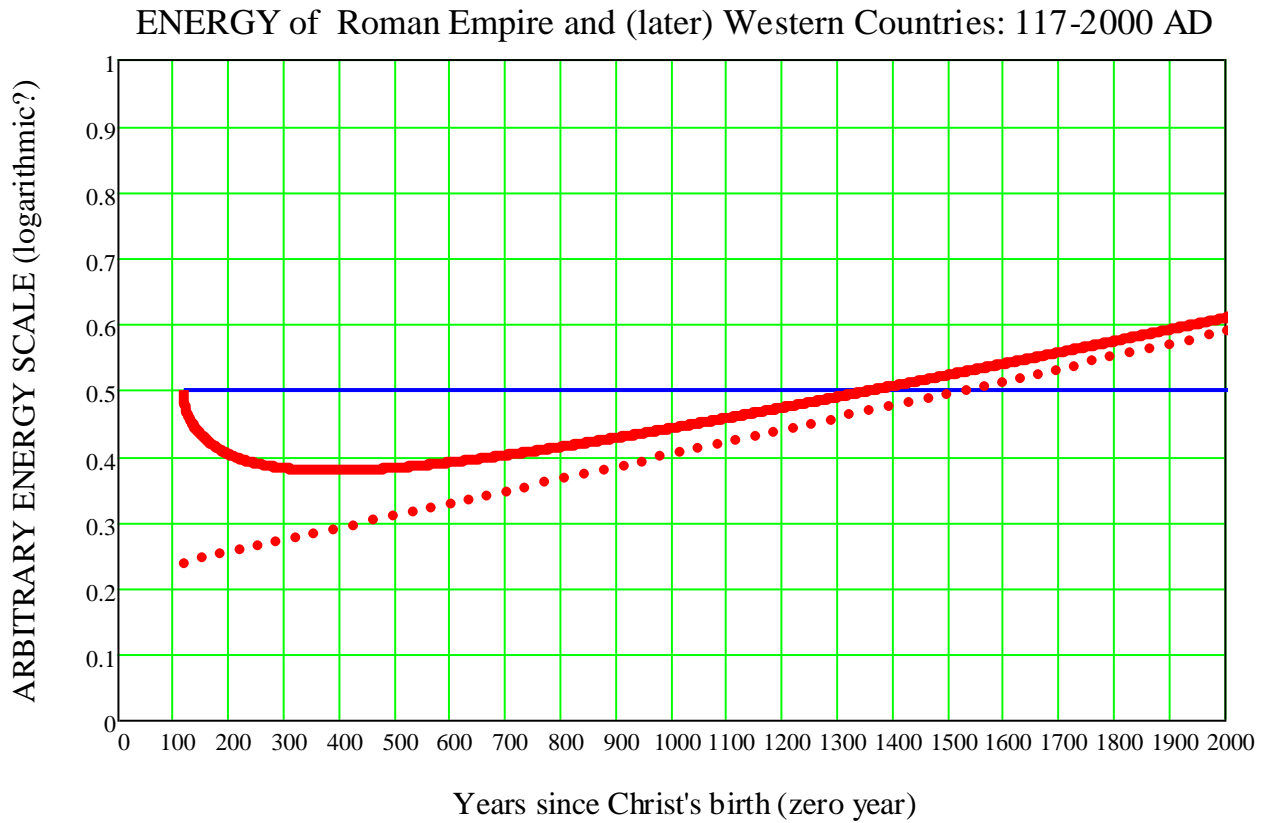

**Figure 7.** Total ENERGY i.e. total WORK produced each year by the Roman Empire starting after its peak, that occurred in the year 117 under Emperor Trajan. This is the solid red curve given by equation (56) of the paper "ENERGY OF EXTRA-TERRESTRIAL CIVILIZATIONS ACCORDING TO EVO-SETI THEORY" (Acta Astronautica, 144 (2018), 202-213). We see that, after Trajan, the empire started to decline, producing less and less total energy and reaching its minimum in the year $385.844 \text{ AD} \approx 386 \text{ AD}$. These were the years (and actually decades, or even a few centuries) of the Barbarian Invasions inside the Western Roman Empire, after the Visigoths had inflicted the first severe defeat to the Romans at the battle of Hadrianople in 378 A.D.. Then, the "Dark Ages of the Western Civilizations", or "Middle Ages", lasted for about ten centuries, and it was not until about 1300 A.D. the Western Europe started flourishing again, reaching about the same Total Energy level that the former Roman Empire had under Trajan. This level is shown in the above graph by the thin solid blue horizontal line. After roughly 1300 A.D., the Italian Renaissance developed and then expanded into the whole of Western Europe in the following centuries. In addition, the dot-dot red line is the oblique asymptote to the Total Energy. Finally, while the horizotal time scale is in agreement with the historic facts, the vertical scale of this graph is completely ARBITRARY, and we had to RE-SCALE it to the correct Energy value (measured in Joules) in our above-mentioned 2018 paper.

## 18. Our two LOGELL HISTORY FORMULAE (discovered in 2018)

In 2019 this author submitted one more paper to the International Journal of Big History, by the title of "TWO POWER CURVES YIELDING THE ENERGY OF A LIFETIME IN EVO-SETI THEORY": the paper was accepted for publication and is currently in press. This paper introduces the new concept of a LOGELL Power Curve, where the part of the curve between peak and death is the descending quarter of an ellipse, rather than a descending parabola, as in the logpar. The name LOGELL actually means "LOGnormal in between birth and peak, and ELLipse in between peak and death". The difference between logpar and logell for the case of Ancient Rome is shown in Figure 8.

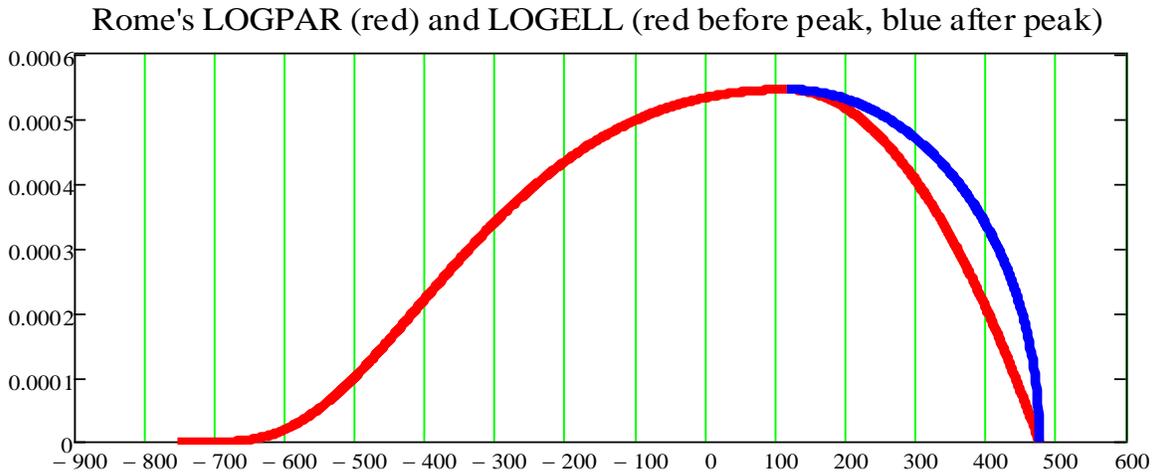

**Figure 8.** The difference between LOGELL and LOGPAR is only in their behaviour between peak $p$ and death $d$ i.e. an ELLIPSE for the LOGELL (in blue above) and a PARABOLA for the LOGPAR (in red above). The common part of the curve prior to the peak (shown in red here) is a b-lognormal, that is a lognormal probability density function (pdf) in the time starting at the birth time $b$ and reaching its peak at time $p$. In this way, the finite lifetime of any living being or civilization is a POWER curve (power means measured in Watts, as in physics) and the area under this power curve is the total ENERGY that the living being or civilization needs in order to cope for its own existence. In fact, the ENERGY is the just integral of the POWER in the time, or, if you prefer, the POWER is just the DERIVATIVE OF THE ENERGY with respect to the time.

Once again it was possible to derive the LOGELL HISTORY FORMULAE, given by:

$$\begin{cases} \sigma = \dfrac{\sqrt{\pi}\sqrt{d-p}}{\sqrt{\pi(d-b)+(\pi-4)(p-b)}} \\ \mu = \ln(p-b) + \sigma^2 \end{cases} \quad (8)$$

And once again $\sigma \to 1$ if we let $d \to \infty$ into the upper equation (8), i.e. "in the long run", as we now explain.

**19.  LOGELL ENERGY approach to IMMORTALITY "in the long run"**

Just as the logpars approach immortality "in the long run", and their respective energies do so by virtue of their own oblique asymptote, so do the logells. Figure 9 hereafter shows both the logpar (in red) and the logell (in blue) asymptotes for Ancient Rome decline, fall and recovery after the Barbarian Invasions. Basically, the difference between logpars and logells is that logells allow for a larger amount of energy (area under each power curve) than the logpars *towards the end of life*. So logells also allow for a *faster recovery* than logpars do. This feature shows neatly up in the two recovery times shown in Figure 9:

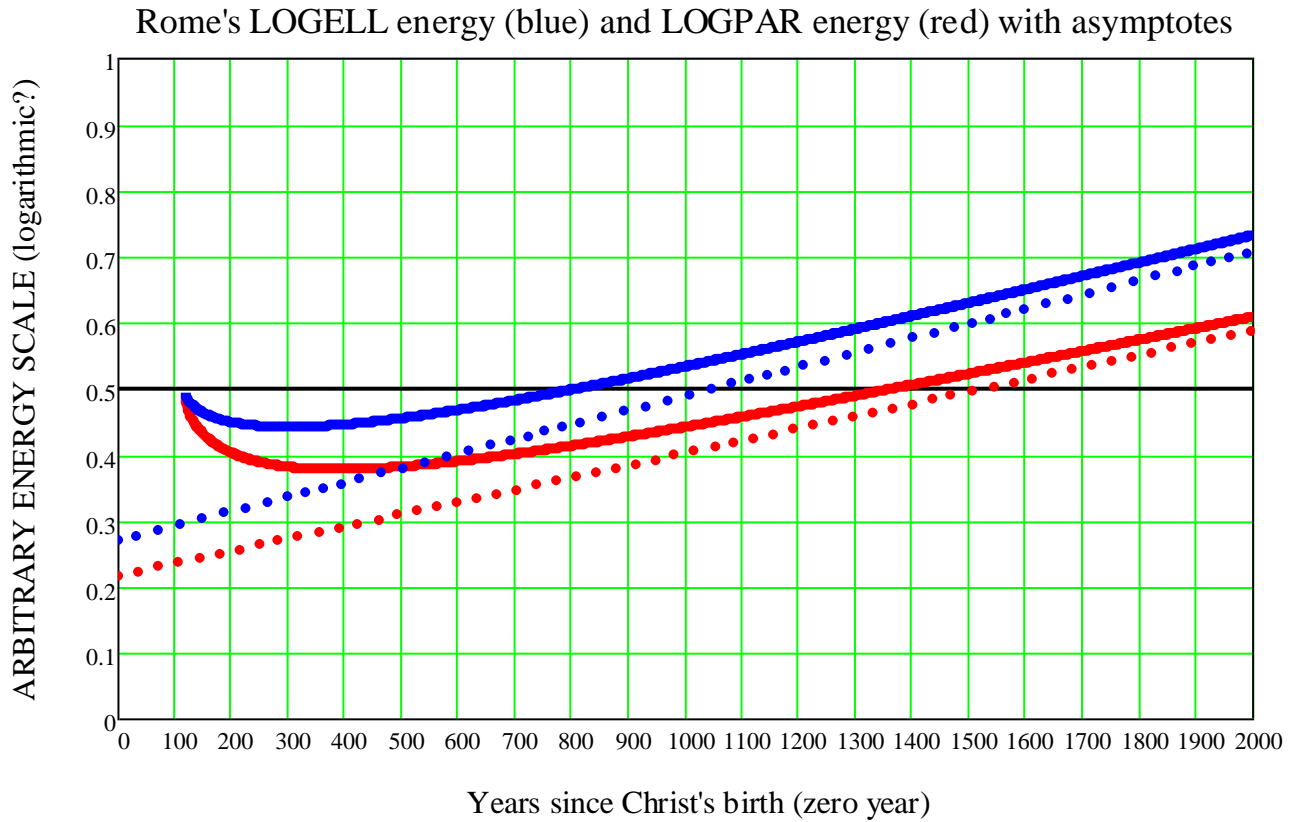

**Figure 9. LOGELL vs. LOGPAR for the RECOVERY of the Western European Civilization: 800 AD or 1400 AD ?**
1) Rome's total energy as a function of $D$ (i.e. $d$, the death time regarded now as the independent variable): The solid curve in red is the LOGPAR ENERGY curve. The dot-dot straight line in red is its oblique asymptote.
2) The solid curve in blue is the LOGELL ENERGY curve. The dot-dot straight line in blue is its oblique asymptote.
3) As we see from Figure 9, the RECOVERY of the Western Civilization after the fall of the Western Roman Empire (476 AD) happened at two different times given by the blue and the red solid curve, respectively. The blue curve (LOGELL) intercepts the 0.5 axis near the year 800 AD, that is the time when Charlemagne was appointed "Holy Roman Emperor" by Pope Leo III and the Carolingian Renaissance started: see the Wikipedia website https://en.wikipedia.org/wiki/Carolingian_Renaissance. The red solid curve (LOGPAR) is the Italian Renaissance https://en.wikipedia.org/wiki/Italian_Renaissance, crossing the 0.5 axis around 1400 AD.

## 20. BIG HISTORY: a new POPULAR way of describing the evolution of the Universe over about 13.8 billion years: Big Bang to Humans (and possibly beyond, into the future, including ETs)

**BIG HISTORY** (https://en.wikipedia.org/wiki/Big_History) is a new research discipline that, in this author's view, is best described by the very same words of the mentioned Wikipedia relevant site. We now report these words just the same. "Big History is an academic discipline which examines history from the Big Bang to the present. Big History resists specialization, and searches for universal patterns or trends. It examines long time frames using a multidisciplinary approach based on combining numerous disciplines from science and the humanities, and explores human existence in the context of this bigger picture. It integrates studies of the cosmos, Earth, life, and humanity using empirical evidence to explore cause-and-effect relations, and is taught at universities and primary and secondary schools often using web-based interactive presentations."

This author became enamoured with Big History in 2013, while he was already working to his mathematical Evo-SETI Theory. He then joined the International Big History Association (IBHA) and attended the relevant Conferences in San Rafael (California, 2014), Amsterdam (The Netherlands, 2016) and at Villanova University (USA, 2018).

Then, he felt confident enough to run his own "Big History and SETI" Conference in Milan, Italy, on July 15-16, 2019, with about 30 speakers from many countries worldwide. All talks were tape-recorded and may be found at the International Big History Association (IBHA) website https://bighistory.org/seti-and-big-history/.

In particular, this author's own presentation of his Evo-SETI Theory is found at the web site: https://www.youtube.com/watch?v=Mcsj_IGuUQU&list=PL2DlytCDrRCnSglJ9_nCbEFWhFgTgYmR-&index=7&t=182s. That video describes all the most important Evo-SETI results that we have mentioned so far.

## 21. Our "E PLURIBUS UNUM" Theorem: i.e. the CONNECTION BETWEEN THE LIFETIME OF MANY SINGLE INDIVIDUALS AND THE LIFETIME OF THEIR CIVILIZATION

We also discovered the mathematical connection (called by us "E Pluribus Unum" Theorem, i.e. "One [Civilization] out of many [Individuals]) between the lifetimes of many single individuals (actually an infinity of single individuals) and the lifetime of the WHOLE Civilization. This connection is the Evo-SETI translation of the Central Limit Theorem (CLT) of Statistics described at the site https://en.wikipedia.org/wiki/Central_limit_theorem.

These results are mathematically rather difficult and were published in the first part of the paper entitled "New Evo-SETI results about civilizations and molecular clock", International Journal of Astrobiology, 16 (1): 40-59 (2017).

## 22. IMPORTANT: the Lifetime in between Birth and Peak ALWAYS IS A b-LOGNORMAL, no matter whether we use a straight line in between senility and death (i.e. a b-lognormal), or a parabola (i.e. a logpar) or an ellipse (i.e. a logell) POWER CURVE for the DECLINE in one's lifetime

The following "obvious" remark is in reality very important: whether we use a b-lognormal, or a logpar, or a logell curve to represent the lifetime of a living being or a civilization, *the part between birth and peak is THE SAME for all curves, and that is just a b-lognormal climbing from birth to peak.*

In other words still: only the second part of the curve, i.e. the one between peak and death, changes according to the assumed decline profile, and is:
1) b-lognormal from peak $p$ + straight line at senility $s$, or
2) logpar descending parabola from peak $p$ to death $d$, or
3) logell descending quarter-of-ellipse from peak $p$ to death $d$.

On the contrary, the area under the curve in between birth and peak is always the same, i.e. the area under the same b-lognormal between birth and peak. We discovered that this area is given by:

$$\text{Birth\_to\_Peak\_ENERGY} = \int_b^p \text{b-lognormal}(t; \mu, \sigma, b)\, dt = \frac{1 - erf\left(\frac{\sigma}{\sqrt{2}}\right)}{2} \quad \text{with} \quad erf(x) = \frac{2}{\sqrt{\pi}} \int_0^x e^{-t^2} dt. \tag{9}$$

In (9) $erf(x)$ is the well-know "Error Function" of Statistics and Probability, i.e. the "integral of the Gaussian".

Note that (9) is independent of both $b$ and $p$ and depends on $\sigma$ only, i.e. we may shift the b-lognormal forth and back along the time axis as we please, but the only variable "that matters" is $\sigma$, i.e. the standard deviation of the original Gaussian now reverberated into the b-lognormal since lognormal=exp(Gaussian).

### 23. ONTOGENY (ONTOGENESIS) and other applications of Evo-SETI Theory to the Bio-Medical Sciences

Ontogeny (until recently also called Ontogenesis, website https://en.wikipedia.org/wiki/Ontogeny) is the part of a baby's life in between birth and puberty. Thus, ontogeny is the "building up" period in any living creature's life: the body "automatically piles up" all necessary and sufficient "tools" to be later transmitted to any offspring.

Around 2018-19, this author discovered an equation giving the ENERGY of ontogeny, but he didn't have the time to publish it, constrained between his duties in other fields of science (astronautics). So, the result that we now show is unpublished, and is mentioned here in public for the first time and without any mathematical proof. With apologies for the lack of proof.

First of all, consider Man as an example. Suppose that the average life of a Man lasts 80 years (we do not make any distinction between men and women at this point). Suppose also that the peak age in a Man's Lifetime is 40. Then, upon setting $b = 0$, $p = 40$ and $d = 80$ into the logpar history formulae (7), they produce the logpar power curve shown in the following Figure 10.

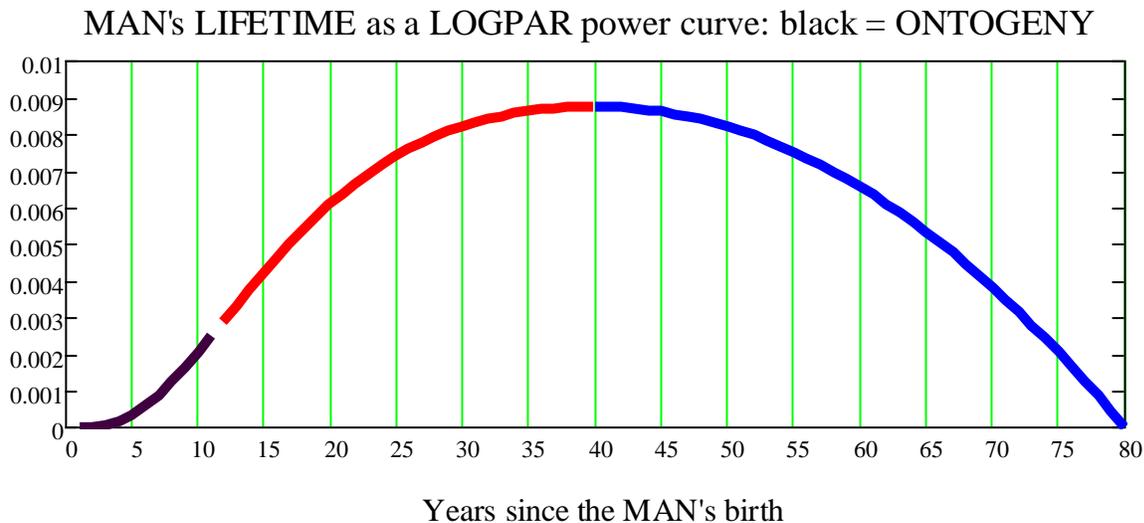

**Figure 10. LOGPAR POWER CURVE of a Man's Lifetime, assuming peak at age 40 and death at age 80 years.**
Three different segments of the logpar power curve exist:
1) ONTOGENY (i.e. Ontogenesis) in between birth and puberty at age around 12 or 13 (BLACK solid curve). Puberty is assumed by us to coincide with the ascending inflexion of the b-lognormal).
2) YOUTH between puberty at 12 or 13 and peak at 40 (RED solid curve).
3) DECLINE between the peak at age 40 and death at 80 (decline's parabola).

The ENERGY OF ONTOGENY is clearly given by the area under the black segment of this logpar power curve. In other words, the energy of ontogeny is given by the integral, that we were able to compute analytically

$$\text{ENERGY\_of\_ONTOGENY} = \int_{birth}^{puberty} \text{b-lognormal}(t;\mu,\sigma,b)\,dt = \frac{1-erf\left(\frac{\sqrt{\sigma^2+4}+3\sigma}{2^{\frac{3}{2}}}\right)}{2}. \qquad (10)$$

Please spend a minute to realize the importance of this result: we are talking about the ENERGY requested to build up a new Living Being from its birth up to its puberty (that we identify with the b-lognormal ascending inflexion time, called *a*=adolescence in previous papers by this author). A lot of further bio-medical research could and should be done, based on equation (10). Unfortunately, at age 71, this author is unable to pursue further research along these lines, but young researchers should think about equation (10).

Actually, this author was able to do better than just (10): he was able to prove that the PROGRESSIVE ENERGY of a Living Being between birth and any instant *z* before the peak is given by the following equation (11):

Birth_to_any_instant_z_before_Peak__PROGRESSIVE_ENERGY__for_any_Living_Being =

$$= \int_{b}^{z} \text{b-lognormal}(t;\mu,\sigma,b)\,dt = \frac{1-erf\left(\frac{\ln\left(\frac{p-b}{z-b}\right)}{\sqrt{2}\sigma}+\frac{\sigma}{\sqrt{2}}\right)}{2} \quad \text{with} \quad b \leq z \leq p. \qquad (11)$$

We leave to the reader to prove that (10) is the particular case of (11) when the upper limit of the integral is abscissa of the increasing inflexion (puberty). Also, just a glance at (11) shows that (9) is the particular case $z = p$ of (11).

Final delight: when you kiss your baby about 4.3 years old, and see that he/she is so *lively,* well… it's all fault of the third derivative of the b-lognormal (5) equaled to zero! In fact, such an equation has three roots:

$$[t = \%e^{-\sigma\sqrt{\sigma^2+3}-2\sigma^2+\mu}+b, t = \%e^{\sigma\sqrt{\sigma^2+3}-2\sigma^2+\mu}+b, t = \%e^{\mu-2\sigma^2}+b] \qquad (12)$$

That, with reference to Figure 10, have the three numerical values (in years since Man's birth):

$$[t = 4.300541166719219, t = 98.07031362681045, t = 20.536684761303683] \qquad (13)$$

Well, to understand the meaning of (13) just remember that the *curvature radius* of any curve b-lognormal(*t*) is given by

$$\text{radius\_of\_curvature\_of\_b-lognormal}(t) = \frac{\frac{d^2 \text{ b-lognormal}(t)}{dt^2}}{\left[1+\left(\frac{d \text{ b-lognormal}(t)}{dt}\right)^2\right]^{\frac{3}{2}}} \qquad (14)$$

Now we introduce an "unjustified approximation", consisting in supposing that one has

$$\frac{d \text{ b-lognormal}(t)}{dt} = 0 \tag{15}$$

during the ontogeny, i.e. during the time of the black solid curve in Figure 10. While this approximation is certainly not justified, yet it might be "numerically not too wrong", as "rough engineers" usually do when applying (14) to their own "practical problems". In this event, (14) reduces to the much easier equation typical of engineers:

$$\text{radius\_of\_curvature\_of\_b-lognormal}(t) \approx \frac{d^2 \text{ b-lognormal}(t)}{dt^2} \tag{16}$$

In this supposition, it follows that the instants of the maxima and minima of the radius of curvature are approximately given by the zeros of the first derivative of (16), that is

$$\frac{d^3 \text{ b-lognormal}(t)}{dt^3} \approx 0 \tag{17}$$

The zeros of this equation (17) are analytically exactly given by (12) and, numerically for the Man input triplet $[b=0, p=40, d=80]$, these three instants (in years since the Man's birth) are given by (13). ***So, when you caress your baby aged approximately 4.3 years, please realize that he/she is living the maximum time of his/hers ontogeny rate! Wow!*** In other words, and much more scientific words than the previous poetic words referring to babies aged 4.3 years, our Evo-SETI Theory shows that (approximately only, because of the "engineers' assumption" (15)) the maximum Ontogeny activity occurs at the time

$$t_{Max\_Ontogeny\_Activity} = e^{-\sigma\sqrt{\sigma^2+3}-2\sigma^2+\mu} + b = (p-b)e^{-\frac{\sqrt{2}\sqrt{d-p}\sqrt{p+5d-6b}+2(d-p)}{p+d-2b}} + b \tag{18}$$

In (18), $\mu$ and $\sigma$ were of course expressed in terms of the input triplet $(b, p, d)$ by virtue of the Logpar History Formulae (7).

Next: what about the second root in (12), numerically corresponding to 98.070 years? Answer: this root must be discarded since it goes beyond the assumed death age of 80, and so it has no biological meaning at all.

And what about the third root in (12), that is

$$t_{Age\_of\_REASON} = e^{\mu-2\sigma^2} + b = (p-b)e^{-\frac{2(d-p)}{p+d-2b}} + b \tag{19}$$

Well, this is no less than the "Age of Reason" i.e. the time in one's life when he/she decides what to do in the rest of his/her life! And for Man it equals 20.536 years, as (13) shows.

### 24. Letting MAXIMA (NASA symbolic manipulator of the Apollo Flights to the Moon) do the CALCULATIONS

No mathematician, however good, can probably conduct lengthy calculations by hand any more. Since the 1950s, however, excellent "symbolic manipulators" have been created to help mathematicians. As of 2020, Mathematica, Maple and a few other codes can do Algebra, the Calculus, Differential and Integral Equations, and even Tensors without making mistakes that at not programming mistakes.

This author loves Maxima (formerly called Macsyma, see the website http://maxima.sourceforge.net/). In fact, Maxima comes to you *for free*, and that's what the students want. Also, since Maxima was created over 50 years ago at the AI Lab of MIT with NASA funds (to check the orbits to the Moon), after 50 years of use it is bug-free.

Chapter 1of this book is thus the Maxima code proving that Molecular Clock Straight Line in Figure 3 is actually the EvoEntropy (= Shannon Entropy with reversed sign and starting at zero just 3.5 billion years ago) of the Running b-Lognormal (RbL) of the Lognormal Process L(t) starting at t=ts and having an *arbitrarily* assigned mean value m(t).

This is the best result of Evo-SETI Theory, since it may be extended to Exoplanets and SETI.

## 25. CHARGE !

Dear Reader,

Are you convinced that Evo-SETI Theory leads to profound numerical insights touching not just Astrophysics, History, Big History, and Biology, but also Sociology, Psychology and other fields of science?

This author hopes so, especially since just today he is 72 years old, and it is high time for him to pass the testimony on to new generations of mathematically trained scientists of all kinds.

Best wishes!

Claudio Maccone                                         Turin (Torino), Italy, February 6th, 2020.